**Title:** Self-supervised learning for label-free segmentation in cardiac ultrasound

**Authors:** Danielle L. Ferreira PhD[1], Connor Lau[1], Zaynaf Salaymang RDCS[2], Rima Arnaout MD[3*]

[1]Department of Medicine, Division of Cardiology
Bakar Computational Health Sciences Institute
University of California, San Francisco
521 Parnassus Avenue
San Francisco, CA 94143

[2]Department of Medicine, Division of Cardiology
505 Parnassus Avenue
San Francisco, CA 94143

[3]Department of Medicine, Division of Cardiology
Bakar Computational Health Sciences Institute
UCSF-UC Berkeley Joint Program in Computational Precision Health
Department of Radiology, Center for Intelligent Imaging
University of California, San Francisco
521 Parnassus Ave Box 0124
San Francisco, CA 94143
Email: rima.arnaout@ucsf.edu

* Corresponding Author

**Short title:** Self-supervised learning for cardiac ultrasound

**KEYWORDS:** deep learning, machine learning, semantic segmentation, self-supervised learning, cardiovascular imaging, ultrasound.

**WORD COUNT:** 4027

**Abstract**

Segmentation and measurement of cardiac chambers is critical in cardiac ultrasound but is laborious and poorly reproducible. Neural networks can assist, but supervised approaches require the same laborious manual annotations. We built a pipeline for self-supervised (no manual labels) segmentation combining computer vision, clinical domain knowledge, and deep learning. We trained on 450 echocardiograms (93,000 images) and tested on 8,393 echocardiograms (4,476,266 images; mean 61 years, 51% female), using the resulting segmentations to calculate biometrics. We also tested against external images from an additional 10,030 patients with available manual tracings of the left ventricle. $r^2$ between clinically measured and pipeline-predicted measurements were similar to reported inter-clinician variation and comparable to supervised learning across several different measurements ($r^2$ 0.56-0.84). Average accuracy for detecting abnormal chamber size and function was 0.85 (range 0.71-0.97) compared to clinical measurements. A subset of test echocardiograms (n=553) had corresponding cardiac MRIs, where MRI is the gold standard. Correlation between pipeline and MRI measurements was similar to that between clinical echocardiogram and MRI. Finally, the pipeline accurately segments the left ventricle with an average Dice score of 0.89 (95% CI [0.89]) in the external, manually labeled dataset. Our results demonstrate a manual-label free, clinically valid, and highly scalable method for segmentation from ultrasound, a noisy but globally important imaging modality.

## Introduction

Artificial intelligence (AI) has the potential to revolutionize medical imaging. The revolution will not be supervised[1].

Nowhere is the potential for AI, as well as the burden of supervised learning, more clear than for cardiac ultrasound, the primary cardiac imaging modality[2]. The quantification of chamber size, mass, and function are critical to diagnosis, prognosis, and management[3]. However, this quantification is laborious, requiring several manual annotations per exam. Furthermore, even when performed by experts, manual annotations can be susceptible to inter- and intra-observer variability given the low spatial resolution and artifacts inherent to ultrasound imaging[4]. Measurement variability can compound e.g. when linear and area measurements are used to calculate volumes and function[5,6]. Finally, despite the importance of structural and functional measurements for all chambers[3], the right heart and left atrium are often neglected in practice, in large part due to the laborious nature of performing additional annotations.

To overcome these challenges, researchers have turned to deep learning-based semantic segmentation. To date, supervised approaches have been used for this task[7–11], but these require the same manual annotations mentioned above. Therefore, supervised segmentation does not alleviate labeling burden, and instead raises additional concern of using variable and error-prone manual annotations as ground-truth[12]. In fact, to mitigate potential bias from any one person, multiple labelers are advocated, which further increases the cost of labeling[13]; an entire industry has arisen to perform manual labeling[14]. Manual labeling also scales poorly with each additional structure to be labeled, perhaps explaining why most studies of semantic segmentation of the heart to date have focused on the left ventricle[9,15–18]. Emerging foundation models for segmentation of photographic images do not obviate manual labeling (some requiring over a billion manual annotations)[19], require additional manual input at the point of use, and do not work well on ultrasound without additional labor[20].

Self-supervised learning (SSL) has the potential to obviate these problems. Broadly defined, SSL learns to perform a task that conventionally requires supervised learning, but instead uses information generated from the data itself rather than relying on manual human labels. SSL networks are trained with automatically generated labels and human annotation is not required[1,21–2324]. SSL has been used to segment objects in photographic imaging with some success[25], but has been rare in biomedical imaging to date due to the low acceptable margin of error required for such applications[26,27] (Fig. S2). In noisy ultrasound, automated segmentation has proved to be especially challenging[28–30], with some approaches to date reducing, but not obviating, use of manual labels[10,11].

To bridge the gap between supervised segmentation and the challenges of ultrasound imaging, we hypothesized that weak labels could be created using computer vision techniques, circumventing both the time-consuming and subjective nature of human labels. We further hypothesized that using these labels in a self-supervised deep learning pipeline designed to mitigate overfitting and incorporate clinical domain knowledge could segment echocardiograms with performance rivaling clinicians.

## Methods

**Overview.** Echocardiogram images were used to develop a self-supervised pipeline (Fig. 1) for cardiac chamber segmentation of the apical-2chamber (A2C), apical 4-chamber (A4C), and short-axis mid (SAX) views. Initial weak labels derived from computer vision techniques together with aggregate statistical information about chamber shapes and relationships were created (Supplemental Methods). These were used to train the neural networks described below and in Figure 1 in a series of early-learning and self-learning steps to arrive at a final prediction. The pipeline's final segmentation predictions were used to calculate structural and functional measurements according to clinical guidelines.

**Datasets.** *Internal dataset.* A total of 8,843 unique deidentified echocardiograms from UCSF were used (Table 1), with waived consent in compliance with the UCSF IRB. All clinically interpretable studies were included regardless of image quality. Top ten indications for study included arrhythmia/abnormal EKG (13%), heart failure/cardiomyopathy (13%), valve disease/murmur (11%), heart disease not otherwise specified (9%), CAD/chest pain (8%), chemotherapy monitoring (7%), perioperative assessment (6%), dyspnea (5%), hemodynamics (5%), and pulmonary hypertension/right heart failure (5%). A2C, A4C and SAX b-mode views were identified as previously described [31]. Views with color Doppler or LV contrast were excluded. For training and validation datasets, only images with a 200mm field of view (FoV) were included. Only images from the first heart cycle were used to reduce image-level redundancy[32,33]. *Training and validation.* 2,228 videos (A2C, A4C, and SAX; 93,000 images) from 450 echocardiograms were used. Eighty percent were used for training and 20% for validation. *Testing.* 8,393 echocardiograms (4,476,266 images) were used as a holdout test set. Measurements, such as left ventricle (LV) ejection fraction (LVEF), LV end-diastolic volume (LVEDV), LV end-systolic volume (LVESV), LV mass index (LVMI), left atrial (LA) volume, right atrial (RA) volume, and right ventricular end-diastolic area (RVEDA) were extracted from the echocardiogram database and used as ground-truth for performance evaluation for test echocardiograms. Where indicated, measurements were indexed by body surface area. Additionally, for echocardiograms with corresponding CMRs, corresponding measurements were extracted from the clinical CMR reports (thus, CMR-derived measurements were performed as per clinical guidelines for CMR). To evaluate test performance for a given chamber, only studies with available clinical measurements were used (Table 1, Fig. S1). Training, validation, and test sets did not overlap by image, patient, or study.

*External dataset.* An external test dataset was obtained from (https://echonet.github.io/dynamic/). It consisted of 20,060 A4C images from 10,030 patients (one systolic and one diastolic image per patient), with manually annotated clinical tracings of the left ventricle, as well as a computer-estimated LVEF, LVESV, and LVEDV for each patient. Pixel size, alternative views, and manual tracings of LA, RA, and RV were not available.

**Data preprocessing and initial weak label extraction.** The ultrasound region of interest was extracted from DICOM images and normalized to a size of 0.5mm/pixel and resized to 256x256 pixels for segmentation networks or to 480x480 pixels for edge detection networks. Pixel intensities were normalized from 0 to 1. Computer vision and clinical knowledge techniques for initial weak label extraction are detailed in the Supplement. Preprocessing used Python 3.6 libraries OpenCV v4.4(https://opencv.org/), scikit-image (https://scikit-image.org/), Scipy (https://pypi.org/project/scipy/) and NumPy v1.20 (https://numpy.org).

**Neural network architectures.** For each training step in Fig. 1, 80 percent of data was used for training and 20 percent for validation (split by patient). *Segmentation.* UNet is a neural network that has proved robust for segmentation in medical imaging. It was therefore used for segmentation as described[34] with the following modifications: the 1x1 output layer was sigmoid-activated, Adam optimizer was set with a learning rate of 1e-4, soft Dice loss and batch size 32 were used. Data augmentation consisted of random modifications to ~20 percent of training data as follows: rotations from 0-10 degrees, width and height shifts ±10% zoom ±20%, shear 0-0.03, horizontal flips, contrast stretching between 2$^{nd}$ and 98$^{th}$ percentile, cropping of 160x160 pixel patches, downscale from 0.25 to 0.5, pixel dropout of 0.1, median blur, Gaussian noise with zero mean and variance between 0.03 and 0.2 , contrast limited adaptative histogram equalization with upper threshold value for contrast limiting of 0.02, and random tone curve with scale of 0.1. *Edge-detection network.* In contrast to photographic images or drawings, objects in ultrasound images are known to have poor edges. Therefore, a holistically nested edge detection (HED) network was implemented for edge-detection tasks as described[35] with the following modifications: the model was initialized with ImageNet weights and fine-tuned using the same hyperparameters as in Xie et al [36] with a batch size 8. The following data augmentations were randomly applied: rotations 0-10 degrees, width and height shifts ±10%, zoom ±8%, shear 0-0.03, and horizontal flips.

Models were implemented in Keras 2.2.4 (https://keras.io/, GitHub, 2015) with TensorFlow 1.15.2 (https://www.tensorflow.org/) backend and trained in a NVIDIA Tesla M60 GPU with 8GiB of memory.

*Quality control (QC) for segmentations during network training.* At each step in the pipeline (Fig. 1), resultant segmentations were evaluated by shape descriptor analysis, discarding chambers of unreasonable size, eccentricity, or geometric chamber relationships as in Supplemental Methods.

*Early learning correction during network training.* Deep neural networks have been observed to be robust to label noise and first fit the training data with clean labels during the early learning phase, before eventually memorizing the examples with false labels or artifacts [37–41] . We therefore exploited early learning in training the segmentation networks used in this study. During network training, the transition between the early learning and memorization phases was detected by monitoring the soft Dice loss curve for the validation dataset. During network training, the soft Dice loss curve for the validation dataset was monitored in TensorBoard 2.3.0 (www.tensorflow.org). The transition from the transient phase (early learning) to the memorization phase was detected where the bend of the soft Dice loss curve in a shape of elbow is observed i.e., the point of maximal curvature using standard methods (e.g. kneed, pypi.org). Training was stopped at the elbow point (early stopping; number of epochs for each training step indicated in Fig. 1).

*Self-learning during network training.* In several neural network training steps (see Fig. 1), the neural network trained via early learning (with only a small number of examples that passed QC) was then used to infer on *all* available training and validation data in a self-learning manner[42], to recruit additional labeled training examples and higher-quality examples for a second round of training.

**Clinical calculations.** The outputs of the segmentation pipeline were used to compute chamber dimensions, i.e., areas and volumes, for A2C, A4C and SAX views according to clinical guidelines[3] using the biplane method of discs for chamber volumes, and using the area-length method for LV mass, and the standard method for LVEF ([LVEDV-LVESV]/LVEDV) with the following exception: in the external dataset, LVEF was estimated from the A4C view only, as a ratio of pixel areas/pixel lengths.

For each metric, we inferred the segmentation model in all frames of all videos that included the chamber of interest. Systolic and diastolic frames were automatically detected by plotting areas of each chamber segmentation by frame for all videos[43] and a sinusoid with heart rate frequency was fitted to these data. Then, the $r^2$ between the area by frame and its sinusoid fitting was used to select the video with the best periodic function per patient. The frame with the largest area of the chosen video by the $r^2$ method was selected to be the diastole, and the smallest frame of the same video to be the systole. (This was done to avoid videos in which the heart may drift in and out of view, which is not uncommon in clinical practice.) This method failed (a sinusoid could not be fitted) on less than 2% of the entire dataset, in which cases diastole and systole frames were manually chosen.

**Statistical Analysis.** Pearson correlation was used to measure the linear relationship between chamber measurements. Linear regression analyses were performed to measure the strength of these relationships. Bland-Altman (BA) plots were analyzed to demonstrate the bias and 95% limits of agreement (two standard deviations) between different methods for each measurement. Cohen's Kappa was used to assess the agreement between the normal and abnormal values determined by different measurement methods (e.g., clinical vs SSL).

Mann Whitney U testing was performed for patient demographics but not reported because due to dataset size, even minimal differences were found to be statistically significant. For r, $r^2$, and Dice, 95% confidence intervals were computed by bootstrapping 10,000 iterations, however, CI were so narrow that they are not reported. All statistical testing was performed using the scipy package (https://pypi.org/project/scipy/) in Python 3.

**Code Availability.** Code will be made available upon publication at github.com/ArnaoutLabUCSF/CardioML

**Data Availability.** Data is available at https://echonet.github.io/dynamic/

## Results

**Overview.** We developed a pipeline (Fig. 1) to provide self-supervised segmentation of echocardiograms. We first extracted weak labels for cardiac chambers using traditional computer vision techniques and clinical shape priors (Methods, Supplemental Methods). We then used these weak labels to train more accurate segmentations, utilizing early stopping and clinical domain-guided label refinement in successive training steps to achieve final performance. Segmentation predictions from the final step were used to calculate biometrics for each chamber[3], focusing on the most clinically relevant[44] views: the apical 2-chamber (A2C), apical 4-chamber (A4C), and short-axis mid (SAX).

We used 93,000 images from 450 transthoracic echocardiograms for training and validation and 4,476,266 images from 8,393 echocardiograms for testing, across a range of image qualities, pathologies, and patient characteristics (Table 1, “all-comers”). A subset of all-comers (n=553) had corresponding cardiac MRI (CMR) available within 30 days for additional comparison among to CMR, as CMR is considered the gold standard for cardiac measurements (Table 1, “CMR subset”). We also tested our pipeline against an external dataset of A4C images with an additional 10,030 patients.

**Self-supervised pipeline can learn to segment ultrasound.**

After confirming reports[29,45] that chamber measurements derived from computer vision alone correlate poorly with clinical measures (Fig. S2, initial steps in Fig. 2), we developed the supervised learning pipeline (Fig. 1) described above and in the Methods. Figures 2A-2L show examples of segmentation performance through successive steps in the pipeline for all three views, showing that subsequent steps in the pipeline can correct initial errors and learn rather than fail, even on images across a range of qualities, pathologies, and derangements.

To demonstrate incremental impact of each step of the pipeline on prediction performance and generalizability, segmentations from intermediate steps of the pipeline were compared to clinical area measurements on images from the validation dataset (Fig. 2L-O). The $r^2$ on chamber areas ranged from 0.06-0.22 when using initial weak labels compared to 0.53-0.81 using the full pipeline. Bias and LOA similarly improved with successive training steps.

**Pipeline-derived structure and function measurements are comparable to clinical echocardiogram measurements.** Pipeline-derived measurements in the all-comers dataset were compared to clinical echocardiogram measurements (Fig. 3, Table S1, Fig. S3-S4).

*Left ventricle.* Pearson correlations (r) between the AI pipeline and clinical echocardiogram measurements for LV end-diastolic volume (LVEDV), LV end-systolic volume (LVESV), and LV ejection fraction (LVEF) were 0.84, 0.9, and 0.81, respectively (Table S1). $r^2$ for these measurements were 0.70, 0.82, and 0.65; notably, the $r^2$ for LVEF achieved through the self-supervised AI pipeline is better than those reported from supervised learning[8,46] (Fig. 3A-3C, Table S1). Bland-Altman bias±LOA (two standard deviations) for LVEDV, LVESV, and LVEF were 2.8±51mL, 5.3±31mL, and -5.3±14.6%, respectively, similar to clinician variability studies[15,47–49] and consistent with those achieved from supervised learning[8,46] (Fig. 3A-3C, Table S1, Fig. S3). Correlations for LV mass were lower (r=0.75, $r^2$=0.56), but are still similar to reported benchmarks (Fig. 3D).

*Right ventricle.* $r^2$ were 0.69 and 0.71 for RVEDA and RVESA, respectively, indicating a strong correlation[50] between the SSL pipeline and clinical measurements. r's were 0.83 and 0.84 and bias ±LOA was -0.86±5.4 $cm^2$, 1.61±3.9 $cm^2$ (Table S1, Fig. 3F-3G, Fig. S3).

*Atria.* The $r^2$ for left atrial volume was 0.84, showing a very strong correlation[50]. bias ±LOA was -0.14±20mL, consistent with clinical inter-observer bias and with supervised learning performance reported in the literature[8,47] (Fig 3E). Right atrial volume $r^2$ was 0.76 and bias±LOA between SSL and clinical measurements was -2.1±21mL (Fig. 3H).

*Normal vs abnormal.* For further clinical context, the above measurements were indexed to body surface area where applicable and binned as normal or abnormal according to reference guidelines[3,51]. Accuracies and Cohen's kappa values are presented in Figure 3, Table S1 and Figure S4. Accuracy ranged from 0.71 for LV mass to 0.97 for LVEF. Kappa values ranged from 0.54 to 0.79 showing a moderate to substantial agreement[50] between AI pipeline and clinical echocardiogram measurements. LV mass and RVESA were the exceptions, where kappa values were only fair.

Taken together, measurements derived from self-supervised learning performed similarly to clinical inter- and intra-observer variability and to measurements derived from supervised deep learning.

**Self-supervised segmentation and clinical echo correlate similarly to CMR gold standard.**

Compared to all-comers, the CMR subset is younger, more male, and has more LV dysfunction (Table 1). Also, due to the nature of different clinical imaging protocols, CMR studies did not include measurements for all chambers. Furthermore, methods for measurement differ between echo and CMR modalities. Despite this, correlations between SSL-derived measurements and CMR were similar to those of clinical echo measurements and CMR (Fig. 3, Table S1). Comparison of clinical echocardiogram measurements to CMR served as a benchmark. $r^2$ ranged from 0.67-0.76 for LV size and function. Bland-Altman bias±LOAs were -60±96mL, -34±83mL, 2.5±20%, for LVEDV, LVESV, LVEF. These levels of agreement are similar to those reported in the literature[15,52], with echocardiography systematically underestimating LV systolic and diastolic size. LV mass showed poorer correlation with $r^2$ and Bias±LOA of 0.32 and 11±169g. Comparing SSL-derived measurements to those from CMR showed similar to slightly worse correlations and similar limits of agreement (Fig. 3). $r^2$ for LV size and function ranged from 0.60 to 0.72. Bias±LOA for LVEDV, LVESV, LVEF, and LV mass were -57±108mL, -30±94mL, -1.5±22%, and 20±159g, respectively (Fig. 3D, Table S1, Fig. S3). As with the benchmark comparison of clinical echocardiography to CMR, LV mass from the SSL pipeline showed worse performance with $r^2$ =0.32.

When indexed and binarized into normal vs. abnormal values, accuracies and kappas for LVEDV, LVESV, LVEF, and LV mass were the same or slightly better for the SSL pipeline than the clinical benchmark (Fig. 3A-3D, Sig. S4). Moreover, measurements derived from the SSL pipeline are more sensitive for abnormal biometrics (Fig. S4).

**Performance across categorical thresholds and various study characteristics.**

While clinical guidelines do not report categorical thresholds for right heart measurements, left heart measurements can be further divided into normal, mild, moderate, and severe categories. We assessed categorical accuracy of LV measurements, comparing both clinical and SSL-derived measurments to the CMR gold standard where available (Fig S5A-E). As expected, accuracies for binary classifications were higher than categorical accuracies for both clinical and SSL-derived measurements (p=0.008). However, there was no statistically significant difference between clinical and SSL-derived accuracies (p=0.66). We further evaluated measurements by a range of study characteristics including age, gender, race, ethnicity, study quality, cardiac disease, and others (Fig S5F-U). As visualized by Bland-Altman analysis, performance did not differ significantly by these factors.

**AI pipeline performs well on external data.**

The external dataset offered the opportunity to calcuate Dice scores between SSL-predicted segmentation and manual tracings of the left ventricle, as well as comparison of LV function using estimations for LVEF (see Methods).

The average Dice score comparing SSL to manual tracing was 0.89 (95% CI [0.89]), representing good agreement. (For comparison, inter-observer variability in Dice scores on manual annotations of the left ventricle can range from 0.82-0.93[53].) Failures included examples where the external dataset was mislabeled, as well as correct external examples where the model failed (Fig S6).

With no A2C images available, we estimated LVEF using A4C images alone. When binarized by normal vs. abnormal, accuracy for LVEF was 0.79 compared to estimates provided for the external dataset. Notably, the SSL pipeline predicted segmentations for all four chambers, but labels were not available for LA, RV, or RA for performance evaluation.

**Discussion**

Segmentation is a global, critical, and challenging task in ultrasound—exactly the sort of task that deep learning promises to help with. However, the noisy ultrasound modality presents a conundrum: it is recalcitrant to self-supervised learning, and yet supervised learning would require even more burdensome manual labeling. We solve this problem by developing a pipeline for self-supervised[1] segmentation of cardiac chambers from echocardiograms without any manual annotation or prompting, to our knowledge the first achievement of its kind. Self-supervised learning can facilitate the development medical foundation models without overburdening clinicians.

To demonstrate rigor and generalizability[44,54,55], we tested on large internal and external datasets—over 40 times the size of the training dataset. Furthermore, the all-comers dataset represented a full range of clinical characteristics and real-world image qualities—no image was excluded due to quality or pathology—making the fact that SSL pipeline's comparison to clinical measurements all the more impressive. Both internal and external datasets are enriched for cardiac diseases compared to the general population[56–59]. Several clinical measurements are a function of two (LA size) to four (LV function) successful segmentations, again making our pipeline's good agreement with clinical measurements noteworthy. Because the SSL pipeline uses information intrinsic to the image itself (whether the heart featured is normal or not), and because clinical information used was derived from all-comers rather than just normal cases, we see the SSL pipeline able to perform even when hearts may be significantly abnormal, such as those with CHD. We benchmarked our results against available measurements at several levels, including clinical echo measurements, CMR, inter-observer variability reported in the clinical literature, and supervised learning.

The scaling implications for self-supervised segmentation in echocardiography are clear. For our 450 training echocardiograms alone, we estimate (based on timed manual annotations of a small sample) that manually labeling all chambers in all three views would have taken a human 1,664 hours. The fact that both internal and external datasets are missing many segmentations both systematically (e.g., right heart, left atrium) and randomly are testament to the laborious nature of segmentation in clinical practice as well. While manual segmentation scales poorly with each additional chamber, our human-label free pipeline is able to segment all chambers in an image simultaneously: for example, we predicted segments for all four chambers in the external dataset even though it only had left ventricular manual labels. Self-supervised segmentation in ultrasound has the potential to impute segmentations for large datasets, with beat-to-beat and even frame-to-frame granularity as previously demonstrated[43], for both clinical and research use. At the same time, the efficiency and interpretability are greater than other current models[60,61]. Human learners require about $10^2$ echocardiography studies to become proficient[62]. Thus, our training dataset size of 450 studies approximates a human learner's efficiency, while our scalability (automatic measurements on >18,000 test studies) far outstrips human capability. Where humans are inaccurate, variable, and incomplete in their measurements, this capability is an impactful step toward more complete, more reliable cardiac segmentation and measurement, which is a critical and clinically required component of every echocardiogram.

In this manuscript, we focused on measurements we could validate against available clinical measurements from our dataset, but a range of additional measurements are immediately possible. Scaling to segmentations of other anatomic structures, other views, and other types of ultrasounds are also feasible relatively quickly, using the same techniques demonstrated here.

In achieving self-supervised segmentation for echocardiograms, we demonstrate the effect of combining several deep learning techniques with traditional computer vision and with clinical spatial and geometric information. Computer vision preprocessing choices were driven by the physics of ultrasound imaging. We used two types of neural networks—segmentation networks that are known to detect textures, as well as edge-detection networks to strengthen ultrasound's noisy boundaries—to obtain better predictive performance than could be obtained from either one alone. Using a sequence of neural networks with leveraging early stopping and self-learning allowed each successive step in the pipeline benefitted from more data with cleaner labels. Even from a label as weak as a Hough circle for the SAX view, a reasonable LV segmentation was recovered, demonstrating the power of this approach. Finally, we found that using spatial modeling information in our pipeline as clinical domain knowledge improved performance[63].

Despite promising results, the SSL pipeline has also some limitations. While bias±LOA from the pipeline was comparable to clinical and supervised ML benchmarks, an ideal pipeline would have even tighter limits of agreement. From a practical standpoint, manual supervision of a strategically chosen[32] subset of images may improve on the results presented here; however, the purpose of this study was to demonstrate the potential of a completely human-label-free approach.

While not a function of the SSL pipeline itself, selection of image frames to serve as systolic and diastolic timepoints in real-world ultrasound clips is also a potential source of error (and an open problem which affects supervised segmentation as well[8]). If a clinician chose different systolic and diastolic frames for measurement, the final clinical measurement could differ from the AI pipeline even if frame-for-frame segmentations are highly concordant (as shown in the external dataset).

Another limitation not inherent to the SSL pipeline, but rather to the data, is that certain clinical echo measurements such as LV mass and right ventricle, are known to be more variable, contributing to lower observed performance for these measures. Repeated manual measurements for all chambers from multiple observers could better define the gold-standard echo measurement and reduce the potential effect of inter-observer measurement error on the clinical gold-standard, but doing this for thousands of test echocardiograms was not feasible within the scope of this study. Additionally, we look forward to further validation in primary care datasets and/or bespoke patient populations.

In summary, self-supervised segmentation of ultrasound represents a paradigm shift in how, rather than laboring to provide labels for data-hungry machine learning models, we can get machine learning to work for us efficiently, robustly, and scalably even for challenging imaging modalities, in order to solve important problems in cardiology and beyond.

# Tables

## Table 1. Study population demographics.

| | Training & validation set (n=450) | Holdout test set (n=8,393) | | All-comers vs. Training | All-comers vs. CMR subset | Training vs. all-comers vs. CMR subset | External test set (n=10,030) |
|---|---|---|---|---|---|---|---|
| | | All-comers (n=8,393) | Subset with CMR (n=553) | p-value (mwu) | p-value (mwu) | p-value (kruskal) | |
| **Demographics** | | | | | | | |
| Age, years ± s.d.(range) | 58±17 (20-90) | 60±17 (15-120) | 49±16 (19-88) | <0.001 | <0.001 | <0.001 | 68±21 |
| Female, n (%) | 220 (49%) | 3552 (50%) | 193 (40%) | 0.5 | <0.001 | <0.001 | 4885 (48%) |
| White, n (%) | 238 (55%) | 3590 (52%) | 246 (54%) | 0.4 | 0.8 | 0.7 | - |
| Asian, n (%) | 69 (16%) | 1358 (20%) | 63 (14%) | 0.04 | 0.001 | <0.001 | - |
| Latinx, n (%) | 56 (13%) | 850 (12%) | 74 (16%) | 0.8 | 0.03 | 0.09 | - |
| Black or African American, n (%) | 31 (7%) | 567 (8%) | 35 (8%) | 0.4 | 0.6 | 0.6 | - |
| Other, n (%) | 39 (9%) | 541 (8%) | 40 (9%) | 0.5 | 0.6 | 0.7 | - |
| **Machine Manufacturer** | | | | | | | |
| Philips | 335 (74%) | 4826 (69%) | 295 (62%) | 0.009 | 0.002 | <0.001 | 10003 (100%) |
| GE | 60 (13%) | 2029 (29%) | 27 (6%) | <0.001 | <0.001 | <0.001 | - |
| Siemens | 55 (12%) | 182 (3%) | 155 (32%) | <0.001 | <0.001 | <0.001 | - |
| **Clinicians** | | | | | | | |
| Number of Unique Sonographers, n | 26 | 53 | 30 | - | - | - | - |
| Number of Unique Diagnosing Physicians, n | 17 | 34 | 23 | - | - | - | - |
| **Study Quality** | | | | | | | |
| Fair, poor, or technically difficult (%) | 123 (27%) | 2438 (35%) | 64 (13%) | 0.002 | <0.001 | <0.001 | - |
| **Cardiac Disease** | | | | | | | |
| Arrhythmia, n (%) | 127 (28%) | 2062 (29%) | 126 (26%) | 0.6 | 0.2 | 0.4 | - |
| Heart Failure Including Heart Transplant, n (%) | 126 (28%) | 1882 (27%) | 184 (39%) | 0.6 | <0.001 | <0.001 | 2874 (29%) |
| Dilated cardiomyopathy, n (%) | 4 (1%) | 110 (2%) | 14 (3%) | 0.3 | 0.02 | 0.03 | |
| Restrictive cardiomyopathy, n (%) | 1 (0%) | - | 3 (1%) | 0.6 | 0.001 | 0.03 | |
| CAD risk factors & CAD equivalents*, n (%) | 261 (58%) | 4128 (59%) | 228 (48%) | 0.8 | <0.001 | <0.001 | - |
| CAD, MI, cardiac arrest, n (%) | 86 (19%) | 1495 (21%) | 89 (19%) | 0.3 | 0.2 | 0.2 | 2290 (23%) |
| Significant valve disease†, n (%) | 63 (14%) | 1041 (15%) | 62 (13%) | 0.6 | 0.3 | 0.5 | - |
| Significant aortic stenosis†, n (%) | 15 (3%) | 219 (3%) | 11 (2%) | 0.8 | 0.3 | 0.6 | |
| Significant aortic regurgitation†, n (%) | 9 (2%) | 80 (1%) | 18 (4%) | 0.1 | <0.001 | <0.001 | |
| Pericardial disease, n (%) | 27 (6%) | 209 (3%) | 34 (7%) | <0.001 | <0.001 | <0.001 | - |
| Significant pericardial effusion†, n (%) | 7 (2%) | 69 (1%) | 20 (5%) | 0.2 | <0.001 | <0.001 | |
| Aortic disease, n (%) | 5 (1%) | 138 (2%) | 13 (3%) | 0.2 | 0.2 | 0.2 | - |
| Pulmonary HTN, n (%) | 42 (9%) | 531 (8%) | 45 (9%) | 0.2 | 0.1 | 0.1 | - |
| Congenital Heart Disease, n (%) | 28 (6%) | 327 (5%) | 68 (14%) | 0.1 | <0.001 | <0.001 | - |
| Cardiac Prostheses‡, n (%) | 59 (13%) | 1011 (14%) | 78 (16%) | 0.5 | 0.2 | 0.4 | - |
| **Measurements** | | | | | | | |
| BSA, $m^2$, mean ± s.d.(range) | 1.88±0.28 (1.1-2.9) | 1.85±0.26 (0.7-4.3) | 1.89±0.25 (1.2-2.6) | 0.09 | <0.001 | <0.001 | - |
| LV Ejection Fraction, %, mean ± s.d.(range) | 61±12 (8-83) | 59±12 (4-89) | 55±16 (10-83) | 0.02 | <0.001 | <0.001 | 56±13 |
| LV Ejection Fraction <35%, n (%) | 29 (7%) | 444 (6%) | 75 (16%) | 0.9 | <0.001 | <0.001 | 948 (8%) |
| LV Ejection Fraction abnormal (qualitative), n (%) | 79 (18%) | 1454 (21%) | 170 (36%) | 0.1 | <0.001 | <0.001 | - |
| LV Diastology abnormal, n (%) | 281 (65%) | 2551 (38%) | 230 (56%) | 0.2 | 0.01 | 0.02 | - |
| LV End Diastolic Volume Index, $mL/m^2$, mean ± s.d.(range) | 56±23 (19-211) | 54±24 (10-244) | 64±29 (10-234) | 0.09 | <0.001 | <0.001 | 91±46§ |
| LV End Systolic Volume Index, $mL/m^2$, mean ± s.d.(range) | 24±19 (5-172) | 24±19 (3-203) | 32±25 (3-189) | 0.6 | <0.001 | <0.001 | 43±35§ |
| LV Size abnormal, n (%) | 56 (13%) | 929 (14%) | 126 (28%) | 0.6 | <0.001 | <0.001 | - |
| LV Mass Index, $g/m^2$, mean ± s.d.(range) | 91±34 (34-290) | 86±26 (19-238) | 98±30 (48-195) | 0.04 | 0.002 | 0.001 | - |
| LV Mass abnormal, n (%) | 106 (29%) | 1565 (26%) | 127 (34%) | 0.3 | 0.001 | 0.003 | - |
| LA Volume Index, $mL/m^2$, mean ± s.d.(range) | 31±14 (10-98) | 31±13 (8-142) | - | 0.3 | <0.001 | - | - |
| RA Volume, mL, mean ± s.d.(range) | - | 22±13 (4-143) | - | - | - | - | - |
| RV End Diastolic Area, $cm^2$, mean ± s.d.(range) | - | 17±5 (7-40) | - | - | - | - | - |
| RV End Systolic Area, $cm^2$, mean ± s.d.(range) | - | 9±4 (3-28) | - | - | - | - | - |

LV = left ventricle, LV = left atrium, RV = right ventricle, RA= right atrium, BSA = body surface area. mwu, Mann-Whitney U test. kruskal, Kruskal-Wallis test.

*CAD risk factors include hypertension, hyperlipidemia, smoking, family history of CAD, drug use. CAD equivalents include diabetes, peripheral vascular disease, and cerebrovascular disease.

†Significant disease includes any severity greater than mild

‡ includes pacemakers, grafts, balloon pumps, ventricular assist devices.

§ Values are not indexed by BSA.

## Figure Legends

**Figure 1. Overview of the self-supervised segmentation pipeline.** (A) A2C segmentation. (A1) Initial weak label creation using computer vision. (A2) Semantic segmentation training includes a first training with early learning and self-learning to arrive at a second and final trained model used to run inference on test data. (B) A4C segmentation. (B1) Initial weak label creation from inference using the A2C final model from (A), resulting in predictions of two "LA" and two "LV" chambers per image. (B2) Predictions from B1 were reassigned four chamber labels, and clinically-guided morphological operations were performed to stretch the RV apex to known correlations of RV and LV length. Early learning and self-learning were then performed to arrive at the final A4C model. (C) SAX segmentation. (C1) Initial weak labels were generated through Hough circle detection and (C2) a holistic edge detection (HED) network was trained (early learning). (C3) The prediction from C2 was then filled in and used as a label to train a first UNet. After self-learning to recruit more usable labels, dilation and erosion were applied to the UNet prediction to create labels for the epicardial and endocardial areas. A second UNet was trained with these labels, resulting in the final model. (D) Clinical calculations. Predictions from A2, B2, and C3 were used to compute chamber dimensions, areas, volumes, and Dice scores on all-comers and external test datasets, respectively. Number of images, exams, and training epochs specified for each step of the pipeline. LV = left ventricle (red), LA = left atrium (light blue), RV = right ventricle (dark blue), RA= right atrium (pink), HED = holistically nested edge detection, A2C = apical 2-chamber, A4C = apical 4-chamber, SAX = short-axis mid.

**Figure 2. Examples of segmentation and measurement improvement through successive steps of the SSL pipeline.** Evolution of segmentation for several examples for A2C (A-D), A4C (E-H), and SAX (I-L). (A) shows a typical good-quality A2C image: the initial weak label (step A1) segments the chambers poorly but are corrected through successive steps of the pipeline to final prediction (step A2). The pipeline performs well even with left atrial enlargement in a technically difficult image (B), an image where LV contrast (white arrow) obscures the LV lumen (C), and an image with a large LA mass (blue arrowheads) prolapses into the LV (D). (E) shows a typical good-quality A4C image and its segmentation from the initial weak label (step B1) to final prediction (step B2). In (F), a filling defect from an LV thrombus (white arrow) was still present in the final segmentation. In (G), segmentation performs well despite presence of a pericardial effusion (blue arrowheads). In (H), segmentation performs despite RA and RV enlargement and septal flattening due to pulmonary hypertension. (I) shows a typical good-quality SAX image: the initial weak label (step C1) is a simple circle, but segmentations are built up through successive steps of the pipeline to final prediction (step C3). In (J), final segmentation prediction is reasonable despite a slightly off-axis SAX. In (K), segmentation performs despite pulmonary hypertension and septal flattening, but the true degree of septal flattening is blunted in the final prediction. In (L), segmentation performs poorly due to a technically difficult image with dropout in the area of the septum, but still recovers a final prediction despite no prediction in step C2. For each of (A-L), example human segmentations are shown as a visual aid for readers unfamiliar with echocardiograms; note that no manual segmentations were used in training the pipeline. Note too that clinical measurements were not made from a single image frame but rather all image frames in a given cardiac cycle, and often multiple views at once. (M)-(P) show area measurements calculated from segmentations at different steps in the pipeline, using images from the validation dataset. r2 and Bland-Altman bias±LOA comparing these measurements with clinical echocardiogram measurements improve across successive steps in the pipeline. Light blue, limits of agreement (two standard deviations); medium blue, one standard deviation. A2C = apical 2-chamber, LV = left ventricle, LA = left atrium, LOA = limits of agreement.

**Figure 3. Comparison of clinical and SSL-derived chamber measurements** for (A) LV end-diastolic volume (LVEDV), (B) LV end-systolic volume (LVESV), (C) LV ejection fraction (LVEF), (D) LV mass, (E) LA volume, (F) RV end-diastolic area (RVEDA), (G) RV end-systolic area (RVESA), and (H) RA volume. The datasets being compared are listed on the left, along with the number of datapoints in parentheses. Open (white) circles indicate intra-modality comparisons (echocardiography). Pink circles indicate comparisons across modality (to CMR). Light blue bars indicate Bland-Altman limits of agreement (LOA; two standard deviations); darker blue bars indicate one standard deviation. Black asterisk in (D) indicates an assumed Bland-Altman bias, as only LOA was reported. Vertical shading in the kappa comparison plots indicates ranges for poor (purple), fair (blue), moderate (teal), good (green), and excellent (yellow) agreement. Clinical intra-observer references include Jacobs et al 2006 for LVEDV,

LVESV, and LVEF[17]; Crowley et al 2016 for LV mass[64]; and Mihaila et al 2017 for LA volume. Supervised learning references include Ghorbani et al 2020[46] and Zhang et al 2018[8].

**Acknowledgements:** We thank the patients, sonographers and clinicians whose work created the datasets used to develop and test this pipeline. We thank Yumiko Abe-Jones and Nader Najafi for help with data pulls.

**Sources of Funding:** D.F. and R.A. were supported by the National Institutes of Health (R01HL150394) to R.A. R.A. is additionally supported by the Department of Defense (PR181763) and the Chan Zuckerberg Biohub.

**Disclosures:** None relevant.

**Independent data access and analysis:** R.A., D.F., and C.L. had full access to all the data in the study and takes responsibility for its integrity and the data analysis.

**Contributions:** R.A. and D.F. conceived of the study. D.F. and C.L. performed data analysis with input from R.A. Z.S. performed data acquisition with input from R.A. R.A. and D.F. wrote the manuscript with input from Z.S.

**Supplemental Material:** Supplemental Methods; Tables S1–S2; Figure S1-S4

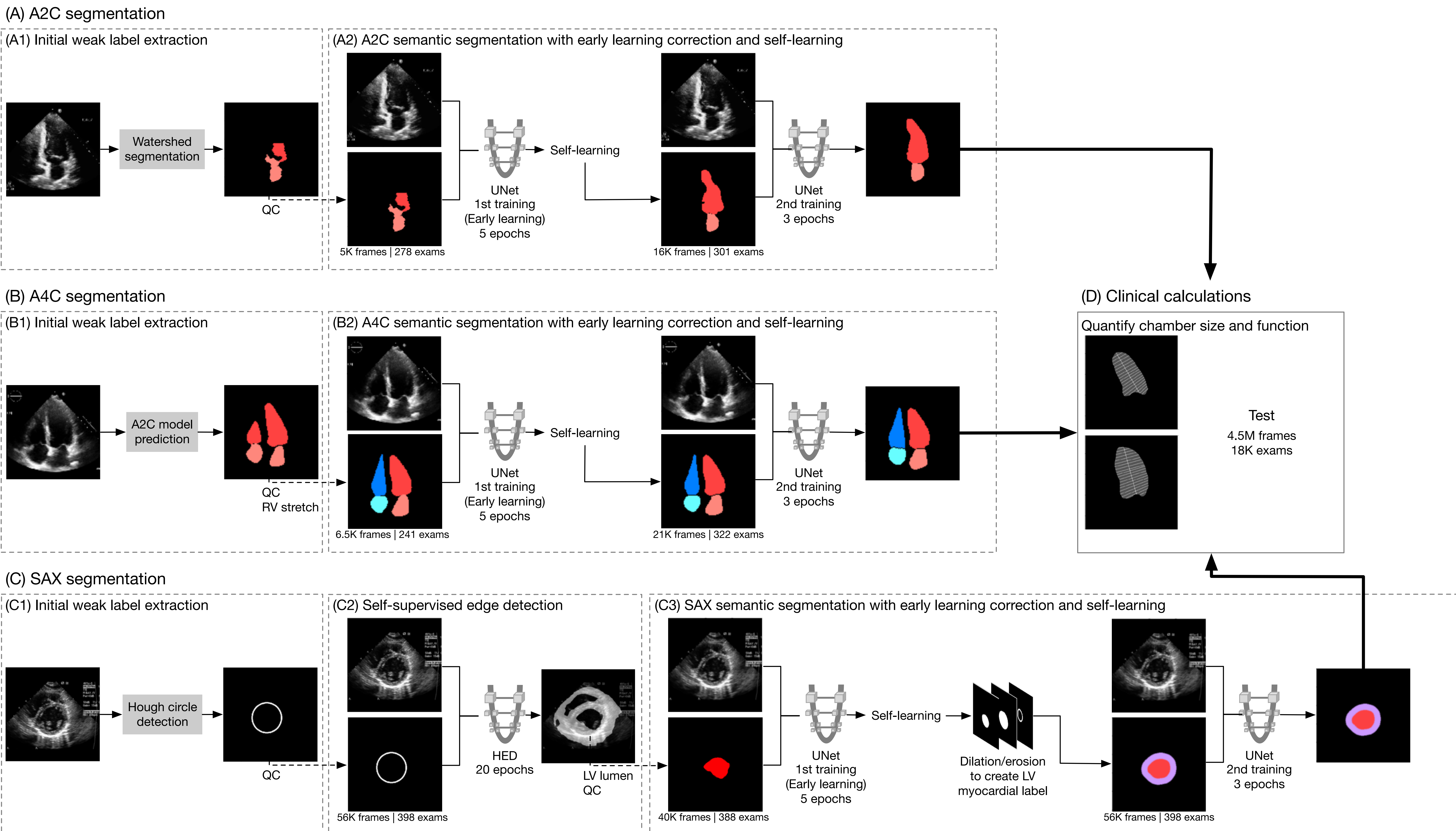

(A) A2C segmentation
(A1) Initial weak label extraction
Watershed segmentation
QC
(A2) A2C semantic segmentation with early learning correction and self-learning
5K frames | 278 exams
UNet
1st training
(Early learning)
5 epochs
Self-learning
16K frames | 301 exams
UNet
2nd training
3 epochs
(B) A4C segmentation
(B1) Initial weak label extraction
A2C model prediction
QC
RV stretch
(B2) A4C semantic segmentation with early learning correction and self-learning
6.5K frames | 241 exams
UNet
1st training
(Early learning)
5 epochs
Self-learning
21K frames | 322 exams
UNet
2nd training
3 epochs
(D) Clinical calculations
Quantify chamber size and function
Test
4.5M frames
18K exams
(C) SAX segmentation
(C1) Initial weak label extraction
Hough circle detection
QC
(C2) Self-supervised edge detection
56K frames | 398 exams
HED
20 epochs
LV lumen
QC
(C3) SAX semantic segmentation with early learning correction and self-learning
40K frames | 388 exams
UNet
1st training
(Early learning)
5 epochs
Self-learning
Dilation/erosion
to create LV
myocardial label
56K frames | 398 exams
UNet
2nd training
3 epochs

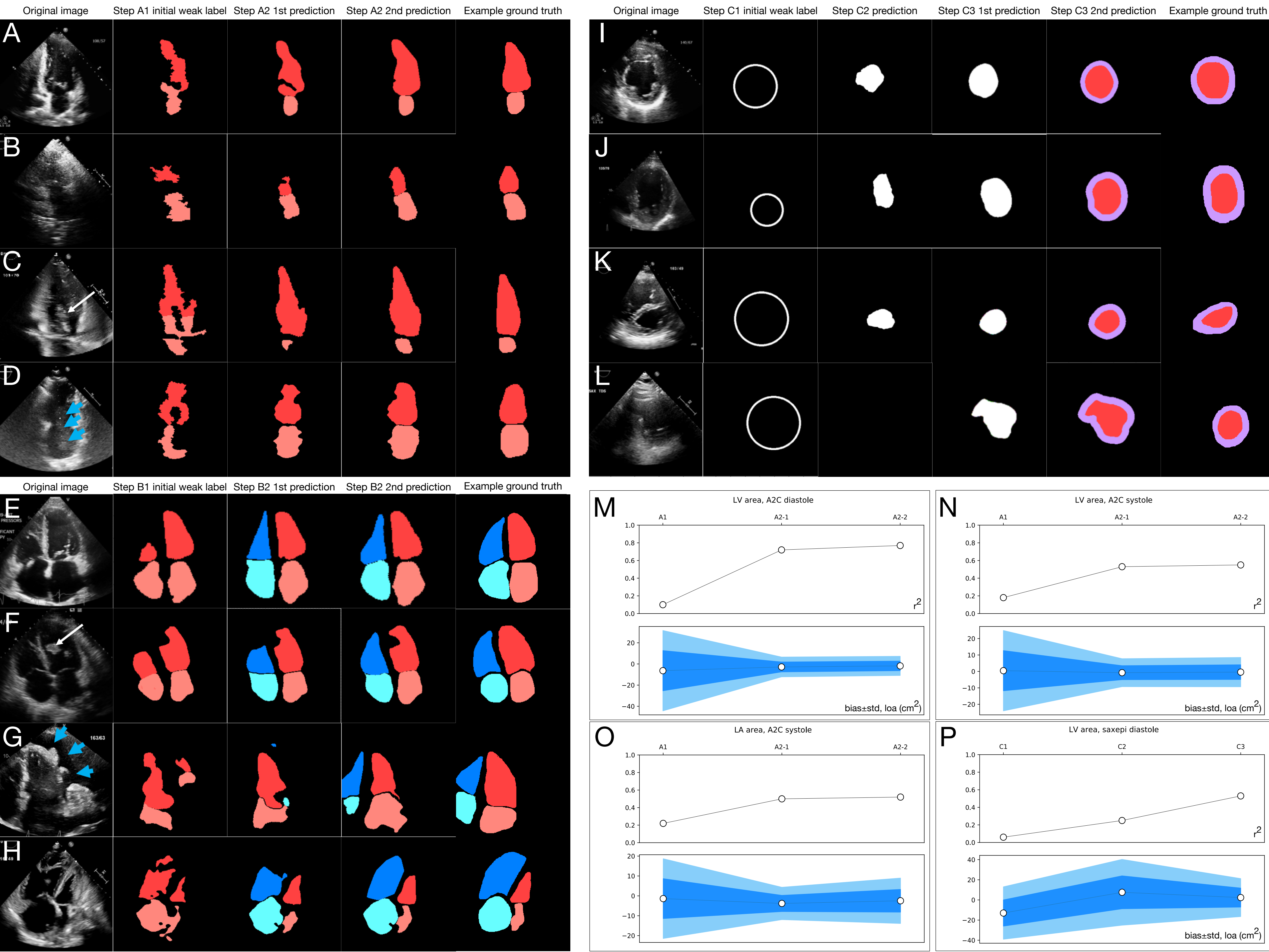

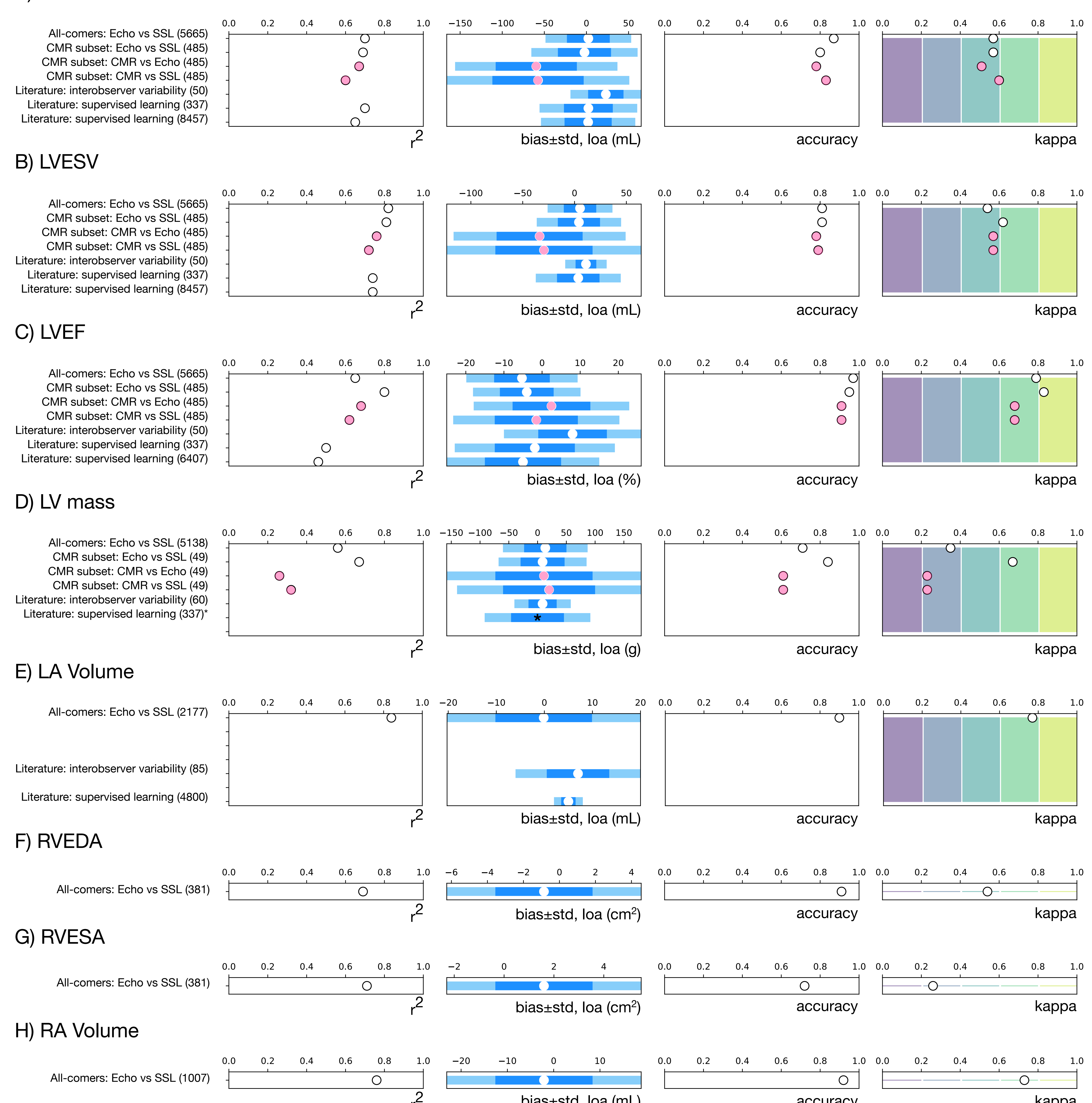

**Title:** Self-supervised learning for label-free segmentation in cardiac ultrasound

**Authors:** Danielle L. Ferreira PhD, Connor Lau, Zaynaf Salaymang RDCS, Rima Arnaout MD

## Supplemental Methods

**Data preprocessing and initial weak label extraction.** *Initial weak label extraction, A2C view.* A bilateral filter ($\sigma_s = 15$, $\sigma_r = 0.25$) was applied to images for speckle noise reduction[1]. A Euclidean distance transform (minimum distance = 20 pixels) was then applied to propose seed points for the watershed algorithm. The watershed algorithm[2] was applied to create initial segments. A copy of this image was binarized (threshold = 0.1) to create a mask for the blood pool, and this mask was applied to remove watershed segments outside the blood pool. Next, connected components analysis excluded images for which the watershed algorithm failed to find at least two segments (since two chambers are expected in the A2C view). The remaining suitable segments were labeled as 'LA' or 'LV' according to the known spatial relationship of these chambers in the A2C view (using clinical knowledge of the A2C view, we expect to see LA in the bottom of the image and the LV above it) . Finally, shape descriptors such as eccentricity and area (see "quality control" below for thresholds used) were calculated, compared with clinical knowledge about plausible chamber shapes and sizes, and were used to eliminate segments not compatible with known anatomy.

*Initial weak label extraction, A4C view.* The final UNet trained for A2C prediction was used to predict segmentations for A4C images (resulting in predictions with two LAs and two LVs per image). Centroids were calculated for each segment and clinical knowledge about known anatomical relationships leveraged to re-label these chambers as RA, RV, LA, and LV. Shape and topological descriptors derived from aggregate clinical knowledge were used to discard images whose segmentations failed relationship, size, and eccentricity thresholds (see "quality control" below), and which lacked four connected components with the expected geometric relationships of the four chambers.

During pipeline training for A4C, intermediate predictions on the validation set were noted to have systematically shorter RV length than appropriate given clinical knowledge about LV-RV length relationships. We suspected this was due to intermediate predictions' failure to correctly understand RV trabeculations. Intermediate predictions were therefore refined based on clinical spatial priors as follows. Length of the predicted LV and RV segments were measured, and its ratio β (RV height/LV height) computed. If the ratio β was less than 0.8, that is RV height was 80% of the LV height value or smaller, the RV segment was stretched to be proportional to the LV.

*Initial weak label extraction, SAX view.* The Hough circle transform[3] was used to leverage clinical knowledge that the LV in the SAX view has a circular shape. A median blur filter (kernel size = 9) was applied to images to reduce noise and a Laplacian operator (kernel size = 5) was employed to extract edges and to reduce the amount of data in the image, an important step to reduce the computation time of the Hough transform. Finally, the Hough circle transform[3] was applied to detect the center and the radius of the endocardial region. Parameters for the Hough transform were a minimum distance between the centers of the detected circles of 400, minimum and maximum circle radius of 20 and 80, respectively; these were chosen empirically based on the approximate size of the LV as a proportion of the image in the SAX view.

For all hyperparameters used in data preprocessing and initial weak label extraction, hyperparameters across a range of 0.1X to 10X of default values were tested on a small sample (n~20) of images. The best parameters chosen by visual review and quantification of what proportion of images had plausible weak labels generated under the various conditions.

**Neural network architectures.** *Quality control for segmentations for initial weak label extraction and during network training.* These quality control thresholds were derived from clinical knowledge about chamber shapes, sizes, and relationships. A2C: Cutoffs for area were 6 to 75cm$^2$ for LA and RA, and 4.7 to 104cm$^2$ for LV and RV. Acceptable eccentricities were 0.3-0.96 for LA, 0.17-0.95 for RA, 0.62-0.96 for LV, and 0.65-0.96 for RV. Connected components analysis required two chambers for the A2C view (an LA and an LV), and four chambers for the A4C view (LA, RA, LV, RV).

## Table S1: Correlation, bias, accuracy, and kappa values.

| Test set | Comparison | N | Regression | | BA analysis | | ICC | Normal vs Abnormal | | | |
|---|---|---|---|---|---|---|---|---|---|---|---|
| | | | r | $r^2$ | Bias | LOA | | Acc | Kappa | F1 | AUC |
| **LVEDV (mL)** | | | | | | | | | | | |
| All-comers | Echo vs AI | 5665 | 0.84 | 0.70 | 2.26mL | ±50.9mL | 0.83 | 0.87 | 0.57 | 0.65 | 0.80 |
| CMR subset | Echo vs SSL | 485 | 0.69 | 0.69 | -2.34mL | ±63.1mL | 0.81 | 0.8 | 0.57 | 0.72 | 0.80 |
| CMR subset | CMR vs SSL | 485 | 0.78 | 0.60 | -57.49mL | ±108.4mL | 0.41 | 0.78 | 0.52 | 0.69 | 0.77 |
| CMR subset | CMR vs Echo | 485 | 0.82 | 0.67 | -59.58mL | ±96.4mL | 0.50 | 0.83 | 0.61 | 0.74 | 0.81 |
| Clinical variability | literature | 50 | -- | -- | 23.0mL | ±42mL | -- | -- | -- | -- | -- |
| Supervised learning | literature | 337 | -- | 0.7 | 2.3mL | ±58mL | -- | -- | -- | -- | -- |
| Supervised learning | literature | 8457 | -- | 0.65 | 2.0mL | ±56mL | -- | -- | -- | -- | -- |
| **LVESV (mL)** | | | | | | | | | | | |
| All-comers | Echo vs SSL | 5665 | 0.9 | 0.82 | 5.3mL | ±31.3mL | 0.88 | 0.81 | 0.54 | 0.66 | 0.83 |
| CMR subset | Echo vs SSL | 485 | 0.9 | 0.81 | 4.24mL | ±40.7mL | 0.88 | 0.81 | 0.62 | 0.79 | 0.83 |
| CMR subset | CMR vs SSL | 485 | 0.85 | 0.72 | -29.5mL | ±93.8mL | 0.60 | 0.79 | 0.57 | 0.79 | 0.79 |
| CMR subset | CMR vs Echo | 485 | 0.87 | 0.76 | -33.7mL | ±83mL | 0.66 | 0.78 | 0.57 | 0.76 | 0.79 |
| Clinical variability | literature | 50 | -- | -- | 11mL | ±20mL | -- | -- | -- | -- | -- |
| Supervised learning | literature | 337 | -- | 0.74 | 3.6mL | ±41mL | -- | -- | -- | -- | -- |
| Supervised learning | literature | 8457 | -- | 0.74 | -- | ±39mL | -- | -- | -- | -- | -- |
| **LVEF (%)** | | | | | | | | | | | |
| All-comers | Echo vs SSL | 5665 | 0.81 | 0.65 | -5.27% | ±14.6% | 0.72 | 0.97 | 0.79 | 0.81 | 0.93 |
| CMR subset | Echo vs SSL | 485 | 0.8 | 0.8 | -4.01% | ±14.1% | 0.86 | 0.95 | 0.83 | 0.86 | 0.94 |
| CMR subset | CMR vs SSL | 485 | 0.79 | 0.62 | -1.51% | ±21.8% | 0.78 | 0.91 | 0.68 | 0.74 | 0.82 |
| CMR subset | CMR vs Echo | 485 | 0.82 | 0.68 | 2.47% | ±20.4% | 0.81 | 0.91 | 0.69 | 0.74 | 0.81 |
| Clinical variability | literature | 50 | -- | -- | 8% | ±18% | -- | -- | -- | -- | -- |
| Supervised learning | literature | 337 | -- | 0.5 | -1.9% | ±21% | -- | -- | -- | -- | -- |
| Supervised learning | literature | 8457 | -- | 0.46 | -5% | ±20% | -- | -- | -- | -- | -- |
| **LV mass (g)** | | | | | | | | | | | |
| All-comers | Echo vs SSL | 5138 | 0.75 | 0.56 | 13.4g | ±73.6g | 0.71 | 0.71 | 0.35 | 0.56 | 0.70 |
| CMR subset | Echo vs SSL | 49 | 0.82 | 0.67 | 8.7g | ±76.5g | 0.83 | 0.86 | 0.71 | 0.86 | 0.86 |
| CMR subset | CMR vs SSL | 49 | 0.57 | 0.32 | 19.96g | ±160g | 0.55 | 0.61 | 0.25 | 0.6 | 0.63 |
| CMR subset | CMR vs Echo | 49 | 0.51 | 0.26 | 11.1g | ±169g | 0.56 | 0.62 | 0.24 | 0.58 | 0.62 |
| Clinical variability | literature | 60 | -- | -- | 8.7g | ±49g | -- | -- | -- | -- | -- |
| Supervised learning | literature | 337 | -- | -- | -- | ±91g | -- | -- | -- | -- | -- |
| **LA volume (mL)** | | | | | | | | | | | |
| All-comers | Echo vs SSL | 2177 | 0.92 | 0.84 | -0.14mL | ±20.1mL | 0.92 | 0.9 | 0.77 | 0.83 | 0.88 |
| Clinical variability | literature | 85 | -- | -- | 7mL | ±13mL | -- | -- | -- | -- | -- |
| Supervised learning | literature | 4800 | -- | -- | 5mL | ±3mL | -- | -- | -- | -- | -- |
| **RVEDA ($cm^2$)** | | | | | | | | | | | |
| All-comers | Echo vs SSL | 381 | 0.83 | 0.69 | -0.86$cm^2$ | ±5.4$cm^2$ | 0.80 | 0.91 | 0.54 | 0.59 | 0.79 |
| **RVESA ($cm^2$)** | | | | | | | | | | | |
| All-comers | Echo vs SSL | 381 | 0.84 | 0.71 | 1.60$cm^2$ | ±3.9$cm^2$ | 0.75 | 0.72 | 0.26 | 0.36 | 0.85 |
| **RA volume (mL)** | | | | | | | | | | | |
| All-comers | Echo vs SSL | 1007 | 0.87 | 0.76 | -2.10mL | ±21mL | 0.87 | 0.92 | 0.73 | 0.78 | 0.86 |

Echo = clinical echocardiogram measurements, SSL = self-supervised AI pipeline-derived measurements, LVEF = left ventricular ejection fraction, LVEDV = left ventricular end-diastolic volume, LVESV = left ventricle end-systolic volume, RVEDA = right ventricular end-diastolic area, RVESA = right ventricular end-systolic area, r = Spearman correlation coefficient, BA = Bland-Altman, LOA = 95% limits of agreement (two standard deviations), ICC = intra-class correlation, F1 = F1 score, AUC = area under the receiver-operator curve. CMR = cardiac MRI. As in main Figure 3, clinical intra-observer references include Jacobs et al 2006 for LVEDV, LVESV, and LVEF[14]; Crowley et al 2016 for LV mass[52]; and Mihaila et al 2017 for LA volume. Supervised learning references include Ghorbani et al 2020[41] and Zhang et al 2018[7].

**Table S2. Model performance of LV segmentation in an external dataset.**

| Dataset | EchoNet dataset (Nb patients) | Average Dice Score, Diastole [95% CI*] | Average Dice Score, Systole [95% CI*] |
| --- | --- | --- | --- |
| EchoNet model † | EchoNet test set (1,277) | 0.927 [0.925 - 0.928] | 0.903 [0.901 - 0.906] |
| Our model | EchoNet test set (1,277) | 0.886 [0.885 - 0.887] | 0.851 [0.850 - 0.853] |
| | Whole dataset (10,030) | 0.886 [0.885 - 0.887] | 0.847 [0.845 - 0.848] |

Nb = number
* Confidence intervals were computed using 10,000 bootstrapped samples
† Values for EchoNet test set were extracted directly from the paper[8] since the EchoNet model is not available.

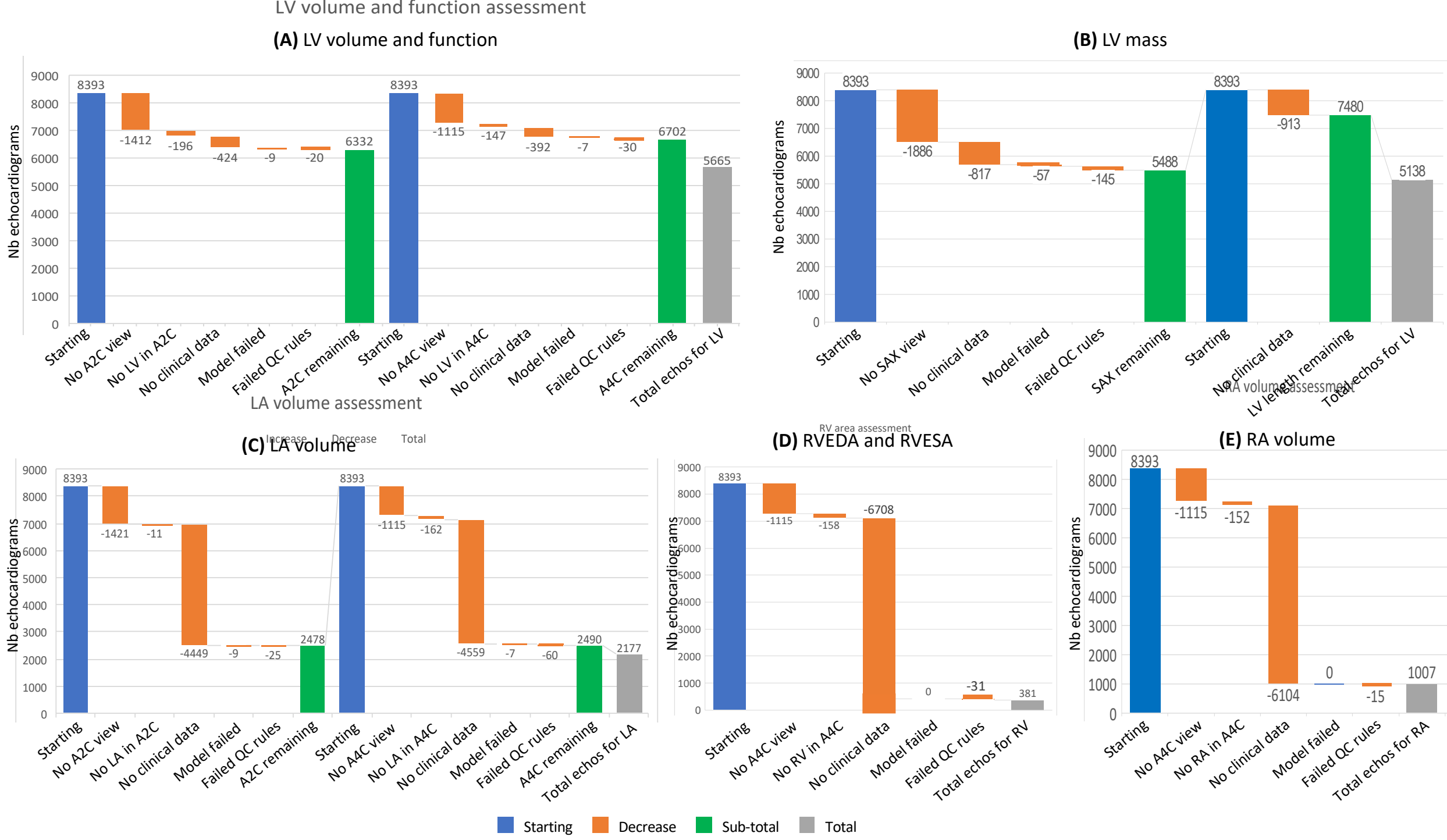

**Figure S1. Inclusion/exclusion of test data by clinical measurement**. The test set had 8,393 echocardiograms/patients. For each measurement assessed, certain views and chambers needed to be present in the echocardiogram as above; a clinical measurement needed to be present in the echocardiogram report for ground-truth comparison. (A) shows the echocardiograms containing A2C and A4C views and clinical ground-truth measurements of the LV for LV size and function assessments. (B) shows the echocardiograms containing SAX views of the LV and LV length measurement for LV mass assessment. (C) shows the echocardiograms containing A2C and A4C views/measurements of the LA, while (D) and (E) show the A4C views/measurements needed for RV and RA assessment, respectively. LV = left ventricle, LA = left atrium, RV = right ventricle, RA = right atrium, QC = quality control, A2C = apical 2-chamber view, A4C = apical 4-chamber view, SAX = short-axis mid view, Nb = number. “No clinical data” = no clinical echocardiogram measurement for comparison to ground truth, “Model failed” = model did not predict anything, “Failed QC rules” = model predicted a chamber, but its shape and/or size failed post-processing rules.

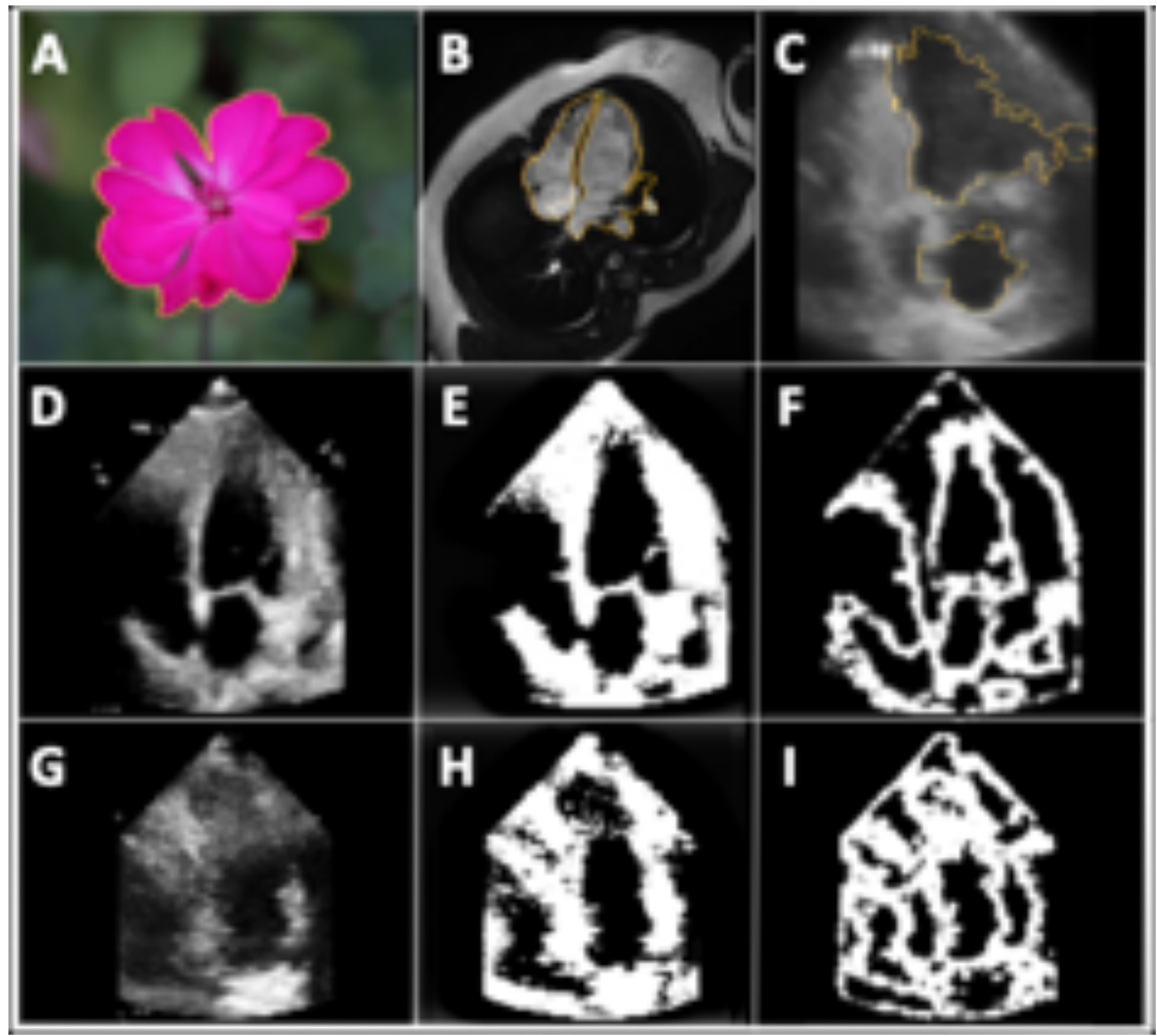

**Figure S2. Examples of standard computer vision methods on different images.** First row – classical computer vision watershed algorithm (thin yellow line) applied to (A) a photo image of a flower, and medical images of the heart in (B) MRI and (C) ultrasound modalities. Note that this algorithm works best on a photo image. Second and third rows depict two echo image examples – left (D,G): original echo images, middle (E,H): segmentation using bilateral filtering, right (F,I): segmentation using optical flow between consecutive frames of the echocardiography sequence.

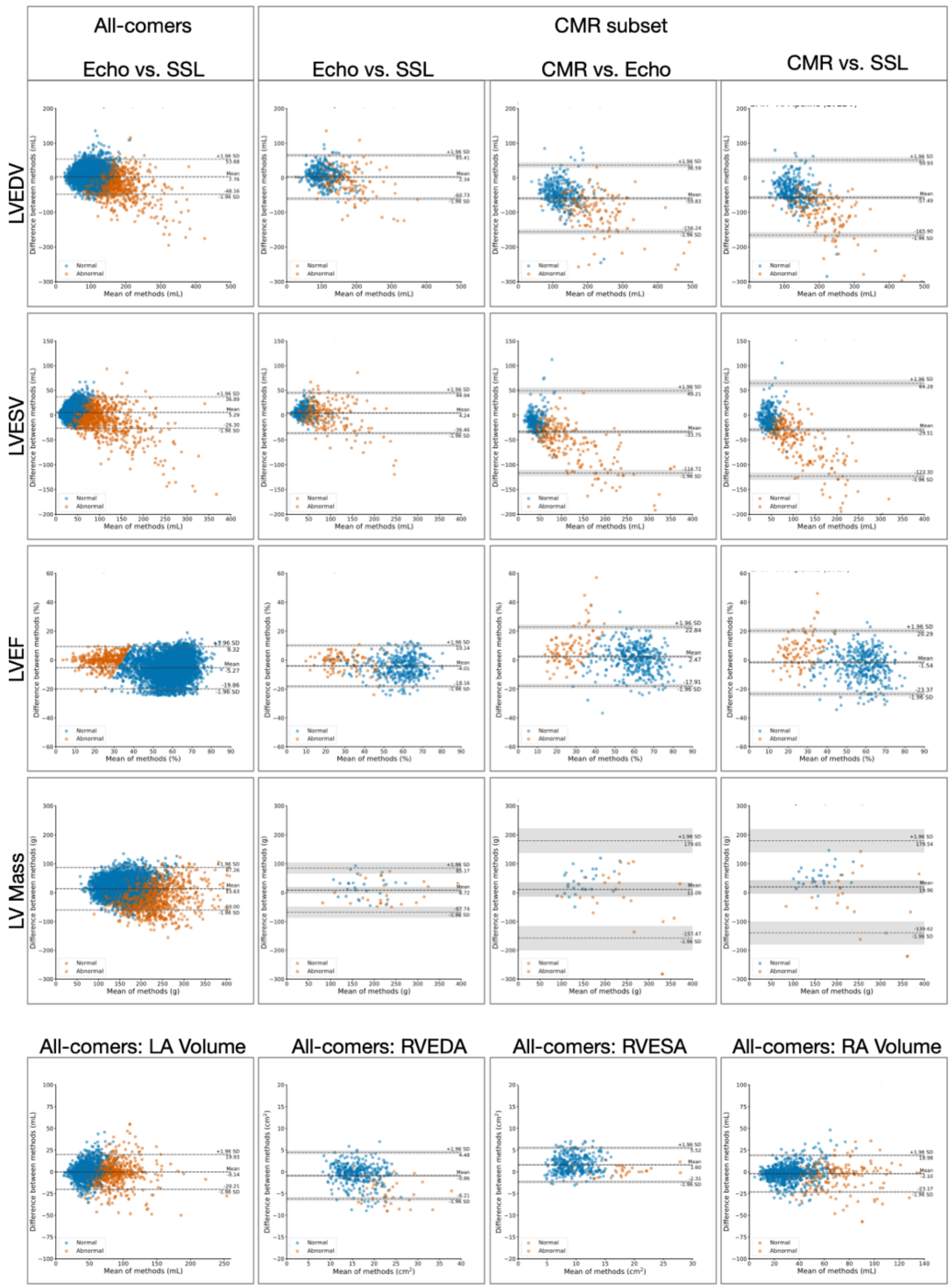

**Fig S3: Bland-Altman plots.** Blue dots represent normal measurements; orange dots, abnormal. SSL = self-supervised AI pipeline, CMR = cardiac MRI, LVEDV= LV diastolic volume, LVESV = LV systolic volume, LVEF = LV ejection fraction, RVEDA and RVESA = RV end-diastolic and -systolic areas, respectively.

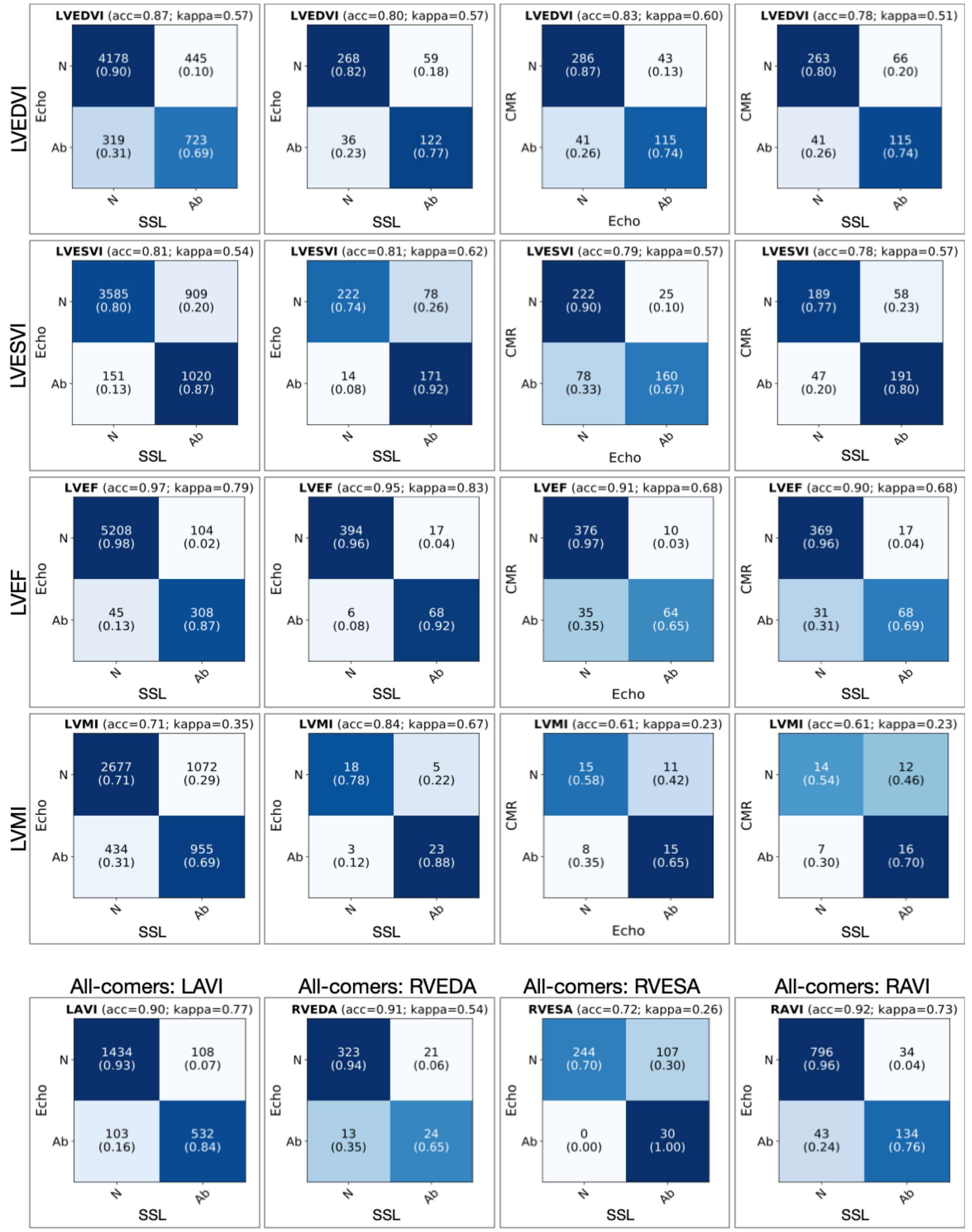

**Fig S4: Confusion matrices for measurements binarized into normal (N) vs. abnormal (Ab).** Accuracy and Cohen's kappa are shown. Measurements are indexed by body surface area where clinically applicable. SSL = self-supervised AI pipeline, CMR = cardiac MRI, LVEDVI = LV diastolic volume index, LVESVI = LV systolic volume index, LVEF = LV ejection fraction, LVMI = LV mass index, RVEDA and RVESA = RV end-diastolic and -systolic areas, respectively, RAVI = RA volume index.

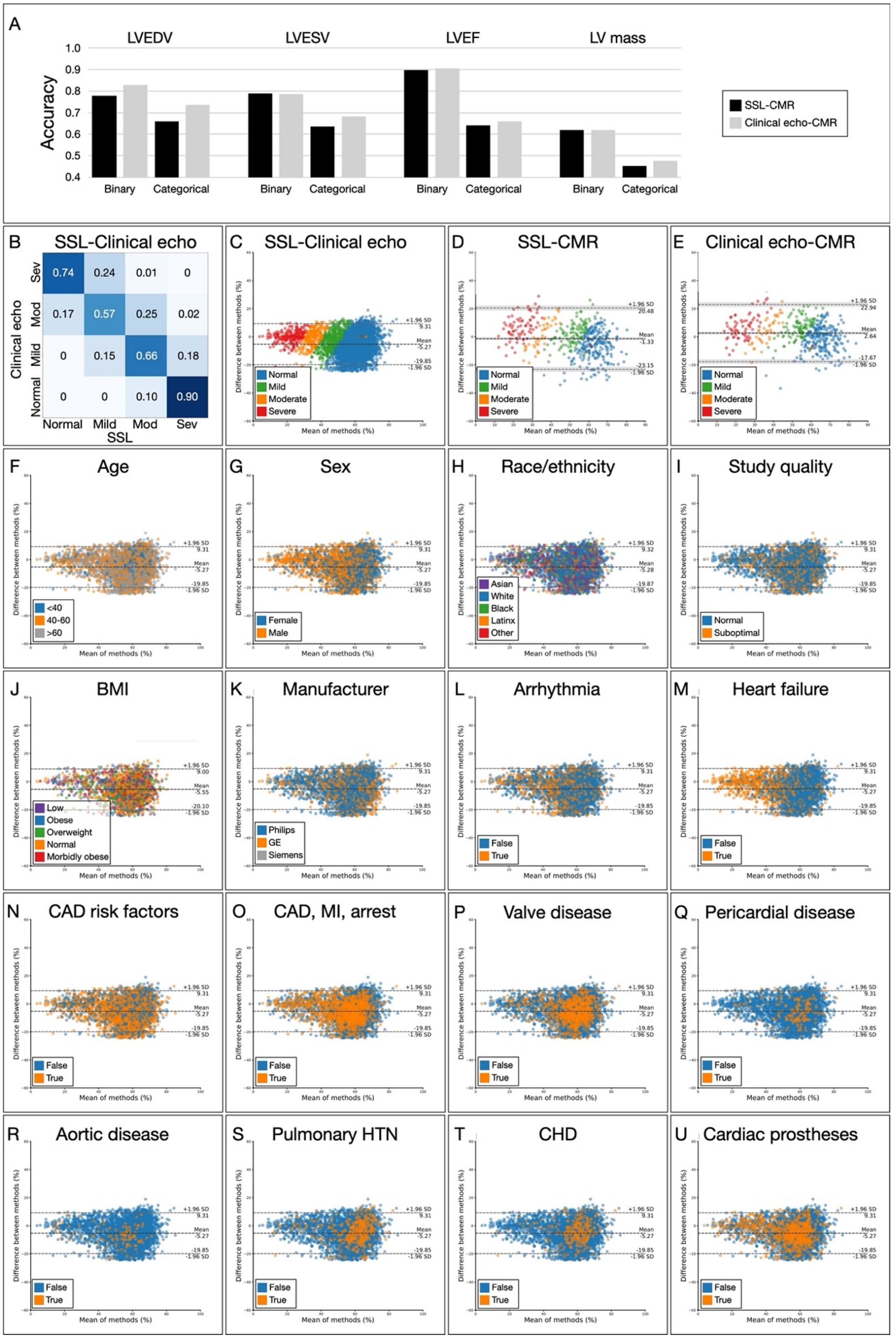

**Figure S5. Performance by categorical values, patient factors, and cardiac pathology.** (A). Accuracy of SSL-derived (black) and clinical (grey) measurements compared to CMR gold standard according to binary breakdown of normal vs abnormal or categorical breakdown by normal, mild, moderate, severe. Accuracies for binary classifications were higher than categorical accuracies for both clinical and SSL-derived measurements (p=0.008). There was no statistically significant difference between clinical and SSL-derived accuracies (p=0.66). Confusion matrix (B) and Bland-Altman plots (C-E) showing categorical performance for LV ejection fraction (LVEF). (C) shows SSL performance compared to clinical echocardiography measurements. (D) and (E) compare SSL and clinical measurements to the CMR gold standard. (F-U) Bland-Altman plots of SSL performance against clinical measurements, colored by patient demographics, study quality, and cardiac pathologies as indicated. LVEDV, left ventricular end-diastolic volume. LVESV, left ventricular end-systolic volume. SSL, self-supervised learning. CMR, cardiac MRI. CAD, coronary artery disease. MI, myocardial infarction. HTN, hypertension. CHD, congenital heart disease. Cardiac prostheses includes pacemakers, grafts, balloon pumps, ventricular assist devices.

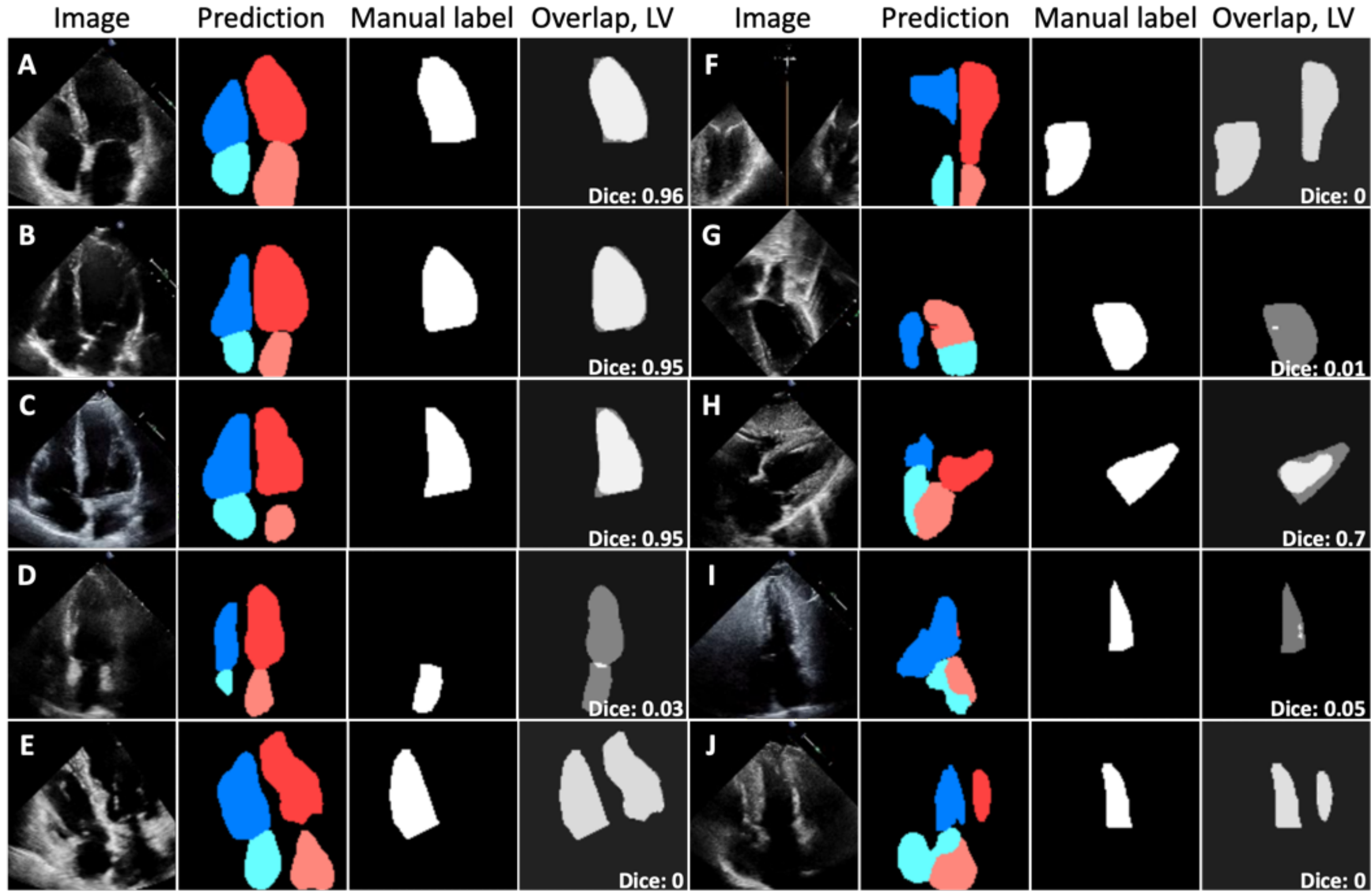

**Figure S6. AI pipeline performance on A4C images from an external dataset: best- and worst-case examples.** This dataset had only A4C images and manually annotated clinical labels for only the left ventricle available. Dice scores ("Overlap, left ventricle") between the AI pipeline's LV segmentation ("AI pipeline prediction," red) and the manual LV segmentation ("EchoNet manual label") are shown. Average Dice score between AI pipeline and manual labels was 0.89 over twenty thousand images. This average Dice includes examples where AI pipeline matched manual label well with high per-image Dice scores (A-C); examples where the AI pipeline was correct but the manual label incorrectly labeled the left atrium (D) or the right ventricle (E) leading to extremely low per-image Dice scores. In these cases, the model was in fact correct despite the low Dice scores. Overall performance also included image view types in the external dataset that that were not part of the AI pipeline training, such as split view (F), inverted view (G), subcostal view (H), and A2C view classified as A4C (I). As these were cases of mislabled views, good model performance was not expected. Finally, it included rare examples where the view and the manual label were correct and the AI pipeline failed (J).

**Supplemental References**